\documentstyle[11pt]{article}  
\begin{document}
\title{Renormalization Theory  of Stochastic Growth} 
\author{Matthew B. Hastings} 
\maketitle
 
An analytical renormalization group treatment is presented of a model which,
for one value of parameters, is equivalent to diffusion limited aggregation.
The fractal dimension of DLA is computed to be $2-1/2+1/5=1.7$.  Higher
multifractal exponents are also calculated and found in agreement with
numerical results.
It may be possible to use this
technique to describe the dielectric breakdown model as well, which is given
by different parameter values.
\section{Introduction and Notation}
\subsection{Introduction}
Diffusion limited aggregation (DLA) is a model for growth of a cluster\cite{Witt}, by the accretion of random walkers.
These random walkers arrive from infinity and stick when they contact the
cluster.  After a walker sticks to the cluster, the next walker is released
from infinity.  This process gives rise to fractal patterns.  
Due to the mathematical equivalence of random walks and potential theory, this
procedure
is equivalent to solving Laplace's equation outside the boundary of
the aggregate, setting the potential zero on the aggregate and constant
at infinity, and picking a point on the surface of the aggregate to
add the walker with a probability proportional to the local field strength;
this field strength may be thought of as an ``electric field".

There has been much numerical work on DLA in two dimensions, where the
fractal dimension has been determined to be 1.71\cite{accepted}.
Mean-field calculations predict $D=5/3$\cite{meanfield}, which
indicates that, in two-dimensions, something is lacking in the mean-field
approach.  In higher dimensions the mean-field theory appears much more
accurate.
One first principles renormalization theory, based on the branching
nature of DLA processes,  obtained 
the result that $D=1.661$\cite{Halsey3}.

Another important analytic result was the derivation of the electrostatic
scaling law, which appears to be obeyed numerically by the aggregates
\cite{Halsey2}.  
This law is used as an essential step in the calculation of this paper.

Recently, another formulation of DLA was proposed, based completely on
conformal mappings\cite{us}.  A conformal function mapped the unit circle
onto the boundary of the aggregate. 
In this formulation of the problem, the electrostatic
scaling law followed almost automatically when considering the behavior
of the first Fourier component of the mapping.
We investigated numerically the problem of the importance of different
Fourier components of the mapping.  It appeared that by directly
simulating the dynamics of only on the first few Fourier components, results
could be obtained for short growth periods that were similar to those when
the full function was used.  This suggested that it might be possible
to develop a renormalization theory based on integrating out higher
Fourier components, using techniques similar to those used in field theory.

Also, by comparing a picture of the cluster generated by only keeping
some small number of terms in the Fourier expansion of the mapping
to a picture generated by the full mapping, it appeared that 
the finite number of terms gave a good description of the boundary of
the object.  It did not accurately describe the exact microscopic
structure of the growth tips, and did not correctly describe the structure
of portions of the object far from the growth region, that is, deep in
the inside of the object.  However, one would expect that the microscopic
structure is not too important, and that the description of regions where there
is little probability of growth is also not important.

Fig.~1 shows a picture of the cluster that results from the conformal mapping
model.  The envelope surrounding the cluster was generated from the first
40 terms of the Fourier series expansion of the 
mapping used to generate the full cluster.

For longer time periods, more terms in the Fourier series were needed, but
this is only to be expected; if only one term were kept in the Fourier series
the object would be a circle and would grow with a radius proportional
to the square root of time.  As more terms are kept, the object can grow
faster than the square root of time by changing shape, but for any given
number of terms, eventually the growth will be as the square root of time.
Therefore, it is expected that in the RG that follows there will be some
cutoff in the number of terms kept which increases with the size of the object.

The above discussion is intended to motivate the RG that follows.  Most
of the discussion is done in more detail in Ref. \cite{us}.

Hope, that such a scheme would work, was provided by numerical evidence
that the conserved quantities (the moments)
of the continuum growth law were very nearly
conserved by the random growth process\cite{moments}.  For the lowest
Fourier coefficients, one would expect that the random growth would
be close to the average growth determined by the continuum law, while
the higher Fourier coefficients would fluctuate more wildly.

The paper is divided into several parts.  First, the conformal mapping model
for DLA is discussed and used to derive continuum equations for growth,
essentially equivalent to the Shraiman-Bensimon equation
for the Hele-Shaw problem\cite{Shrai}.
These equations
are heuristically modified to add the essential differences between DLA
and its continuum limit: the presence of noise and the existence of a 
microscopic cutoff.  This leads to a new model which is hoped to be in the
same universality class as DLA.  Even if it is not in the same
universality class, it is similar enough to be of interest in itself.

Second, under an adiabatic assumption, the equation is modified to
vastly simplify the time dependence, leaving almost a static problem.
The adiabatic assumption makes possible the RG and perturbation theory 
calculations described latter in the paper.
The adiabatic assumption is justified by numerics and 
self-consistently by the RG itself.

At this point, before doing the RG, it is still possible to make
some comparison to numerics based on the continuum representation of DLA.

Third, a perturbation theory is developed for the continuum
equation, with a well
defined set of rules for calculating correlation functions.  The perturbation
theory requires some resummation of diagrams, where to calculate resummed
propagators it is necessary to use a renormalization group approach.
This RG forms the fourth part of the paper; the calculations for the RG
have only been done to lowest order, producing an appropriate renormalized
propagator and vertex.
Fifth, the adiabatic assumption is removed, and the renormalized propagator
is used to calculate various exponents in the theory.
Sixth, the results are compared to numerical experiments, the self-consistency
of the approach is discussed, and there is discussion of what may happen
if the computation is performed to higher orders.
\subsection{Notation}
A large number of functions will need to be defined in this paper.
As much as possible, I will use the following notations.
Capital letters are used for functions, such as $F,G$ to be
defined latter, which describe the shape of a
specific growing aggregate.  Power series expansions of these functions
will be denoted by subscripts, so $F(z)=F_1 z^1 + F_0 z^0 + F_{-1} z^{-1} + ...$
In the continuum limit of these power series expansions, to be appropriately
defined latter, where sums are approximated by integrals, the
Latin letters $j,k,l,m,n,o$ will be used as indices.  
One will see terms like $G(j)$.

Greek letters $\epsilon, \Lambda,\mu$ will be used for 
ultraviolet and infrared cutoffs in these continuum laws.

Greek letters $\alpha,\lambda_0$ will be used for various parameters in
the models defined in this paper.

Lower case letters will be used for functions
which define growth rules for the aggregate.  These include the
functions $f,s,t$ defined latter.  

Latin letters $x,z$ represent points in the complex plane.  The number
$t$ represents time, either as a discrete number of steps or as a real 
number in a continuum limit.  The numbers $\theta,\phi$ represent
angles, while the function $\theta(j)$ is the step function.

Unless otherwise specified, subscripts attached to functions 
will be used to denote derivatives, thus 
$F_x$ is the derivative of $F$ with respect to $x$.  As an exception to
this, the expression $f_{\lambda,\theta}(z)$ will represent
a function parametrized by $\lambda$ and $\theta$.
\section{Conformal Model and Continuum Growth Law}
A model for growth is defined.  From this model, the Shraiman-Bensimon
equation is derived for a function $F$ which maps the unit circle in the
complex plane onto the boundary of the growing object.  Defining 
\begin{equation}
\label{Geq}
G=\frac{1}{F_z}
\end{equation}
and making some approximations, we obtain a continuum growth rule, equation 
(\ref{ceq}), which still includes effects of noise and finite cutoff.

\subsection{Conformal Model}
The following model for DLA in terms of conformal mappings
leads to results that are apparently numerically equivalent to DLA\cite{us}.
It should be noted that the RG in this paper relies upon a continuum 
approximation to this model; this continuum approximation could also have
been obtained from the lattice version of DLA without reference to the
conformal mapping model, but the conformal mapping model provides 
a better justification for the continuum equation.

In this method, we deal only with the analytic function
$F$, which is defined as the analytic function which maps the unit
circle in the complex plane onto the boundary of the growing cluster.
We introduce two parameters, $\alpha$ and $\lambda_0$, where
$\alpha=2$ corresponds to DLA and $\lambda_0$
is some constant determining the size of an individual
random walker.
To grow the object, first pick a random point $x=e^{i\theta}$
on the unit circle.
Then, calculate $F_z(x)$, which is the derivative of $F$ at this point. 
We define 
\begin{equation}
\label{laeq}
\lambda=\lambda_0 (F_z(x) F^*_z(x))^
{-\alpha/2}
\end{equation}

The case of $\alpha=2$ will correspond to DLA and it is that
case that will be considered from now on; other cases will be
briefly discussed in the conclusion.  Then in a given growth step, $F(z)$ is replaced by 
\begin{equation}
\label{grule}
F(f_{\lambda,\theta}(z))
\end{equation}
where $f$ is a function that produces a small bump at angle $\theta$, with
linear dimension of the bump of the order of the square root of $\lambda$.
$\lambda,\theta$ are parameters that define the function $f$.
An explicit example of $f_{\lambda,0}$ is given by:  
\begin{equation}
\label{feq}
\left[\frac{1+\lambda}{2z}(z+1)
\left(z+1+\sqrt{z^2+1-2z \frac{1-\lambda}{1+\lambda}}\right) -1\right]^{1/2}
z^{1/2}
\end{equation}
For $\theta \neq0$, we have 
$f_{\lambda,\theta}(z)=e^{i\theta}f_{\lambda,0}(e^{-i\theta}z)$.

In the small $\lambda$ case, $f$ reduces to  
\begin{equation}
\label{fsl}
z+\lambda z (z+e^{i\theta})/(z-e^{i\theta})
\end{equation}
and by averaging over angle we may determine a continuum growth law.
A picture of a cluster produced by this growth rule is shown in Fig.~1.
A picture of the effect of the mapping $f$ on the unit circle is shown
in Fig.~2.

\subsection{Continuum Growth Law}
The continuum growth law for DLA is known to be equivalent to
the Hele-Shaw dynamics, which obeys the law 
\begin{equation}
{\rm Im}(F_t F^{*}_{\phi})=1
\end{equation}
where $F$ is a function which maps the
unit circle in the complex plane onto the boundary of the aggregate and
subscripts denote derivates with respect to time or to angle $\phi$ on 
the circle.  This law may
be rewritten as 
\begin{equation}
\label{norm}
\begin{array}{l}
 {\rm Re}(F_t/z F_z)=1/(F_z {F}^{*}_z) \\
 {\rm Re}(F_t \frac{|F_z|}{z F_z})=1/|F_z|
\end{array}
\end{equation}
 where here $z=e^{i\phi}$.
Finally, this second growth law may be rewritten as
\begin{equation}
\label{Fceq}
F_t=F_z \int \frac{d\theta}{2\pi} \Bigl(F_z(e^{i\theta}) {F}^{*}_z(
e^{i\theta})\Bigr)^{-1} z \frac{z+e^{i\theta}}{z-e^{i\theta}}
\end{equation}
This equation, the Shraiman-Bensimon equation, results from substituting
the small $\lambda$ expansion of equation (\ref{fsl})
into the equation (\ref{grule}) for the dynamics
of $F$, where $F(f(z))$ is approximated by $F(z)+F_z(z)(f(z)-z)$.

In the Shraiman-Bensimon equation, one may divide both sides by $z F_z$ and
then take the real part of both sides.  This will recover equation
(\ref{norm}) and show that the two equations are equivalent.
Equation (\ref{norm}) implies that the normal velocity of the surface at
a given point is proportional to the local electric field at that point.

Equation (\ref{fsl})
may be rewritten as (taking $\theta=0$ for simplicity)
\begin{equation}
\label{fsle}
z+\lambda z (z+1)/(z-1)=z+\lambda z(1+2/z+2/z^2+2/z^3. . .)
\end{equation}
Therefore, the effect of the integration over 
angle in the continuum growth law is to
project out negative Fourier components in $\lambda$ considered
as a function of angle.  The factor of 2
difference between the zeroth component and all other components will be
important later.

It will also be useful to define a continuum law for another function
$G$ which is defined by equation (\ref{Geq}).  
This function has several advantages.
The equation for $\lambda$ then becomes 
\begin{equation}
\lambda=\lambda_0 G(x)G^*(x)
\end{equation}
which has a simpler form than equation (\ref{laeq}).
This has a physical interpretation that $G$ 
determines the strength of the electric field at point $x$.
Also, $G$ is the derivative of the inverse function of $F$ and the inverse
function of $F$ has a more natural growth rule than $F$ does.  That is,
if $F^{-1}(F(z))=z$, then  under a growth step with
given $\lambda$ and $x$, the function $F^{-1}(z)$ changes into 
$f^{-1}(F^{-1}(z))$.
The continuum law for $G$, as obtained by using the definition of
$G$ and the growth law for $F$,  is 
\begin{equation}
\label{Gceq}
\begin{array}{lll}
G_t&=&G_z \int \frac{d\theta}{2\pi} G(e^{i\theta}) G^{*}(e^{i\theta})
 z (z+e^{i\theta})/
(z-e^{i\theta})\\&& - G \int \frac{d\theta}{2\pi} G(e^{i\theta}) G^{*}(e{i\theta})
[z(z+e^{i\theta})/(z-e^{i\theta}]_z
\end{array}
\end{equation}

It is also useful to consider the continuum growth laws for the power
series of $F$ and $G(z)$.  Writing 
\begin{equation}
\label{FPS}
F(z)=F_1 z^1 +F_0 z^0 +F_{-1} z^{-1} + ...
\end{equation}
\begin{equation}
\label{GPS}
G(z)=G_0 z^0 +G_{-1} z^{-1} + G_{-2} z^{-2} + ...
\end{equation}
then equation (\ref{Gceq}) is equivalent to
\begin{equation}
\label{GPceq}
(G_{-j})_t=(j-2k-1)\sum_{k,l,m} G_{-k} G_{-l} G_{-m}^* \delta(k+l-m-j) 2\theta(j-k)
\end{equation}
where the discrete step function is defined by
\begin{equation}
\label{discretetheta}
\theta(j-k)=
\left\{
\begin{array}{ll}
1 & {\rm for} \; j>k \\
1/2 & {\rm for} \; j=k \\
0 & {\rm for} \; j<k \\
\end{array}
\right.
\end{equation}
An continuum equation may also be written for the power series expansion
of $F$, but we will not need to use such an equation.

There are some problems with directly applying the continuum growth
law above, in any of its forms,
to the discrete random process that defines DLA.  The
continuum law leads to the appearance of cusps in the contour of the cluster
after a finite time, and the continuum law is deterministic while the
discrete law is random.  However, the continuum law must have some
applicability to the discrete cluster growth, because, for example, the
conserved quantities of the continuum law are approximately conserved
by the random process\cite{moments}.  Thus, we will try, in the rest of
this section, to correct the problems in the continuum law so that
it may be of some use in describing the discrete, random growth process.
\subsection{Ultraviolet Cutoff}
The above formulation of the problem suggests a method, outlined
in this section, of inserting
an ultraviolet cutoff into the growth law.
This cutoff will be inserted by hand, and then the parameter of the
cutoff will be adjusted to obtain the correct microscopic scale.

In the discrete conformal mapping model, $F$ never develops cusps
because $f$ is always well behaved.  The specific form of $f$ does
not matter; all that matters is that the approximate expansion for
$f$ given by  $f(x)=z+\lambda z(1+2/z+2/z^2+. . .)$
is only correct for small negative powers of $z$.  The power series expansion
of $f$ is cutoff at some point because $f$ is well behaved.  This cutoff
depends on $\lambda$, which itself depends on the angle $\theta$ at
which growth is taking place.  
The approximation made in inserting the ultraviolet
cutoff into the continuum law
is that {\it the cutoff in the power series expansion for $f$
depends only the  average value of $\lambda$ over the circle at the
time of a given growth step, and not on the exact value of $\lambda$ where
the growth is occurring, where the dependence on the average value of
$\lambda$ is chosen in such a way as to produce the correct
microscopic scale in the DLA growth process}.

Then a simplification follows.  Suppose the regularized form of $f$ is chosen 
to be 
\begin{equation}
f(z)_{\lambda,\theta}=z+z \lambda
\frac{(1+\epsilon)z+e^{i\theta}}{(1+\epsilon)z-e^{i\theta}}
\end{equation}
where $\epsilon$ is a fixed function of  
$\langle \lambda \rangle $, which is 
defined to be the average value of $\lambda$ over
the entire circle.

It is worthwhile also to define
\begin{equation}
\Lambda = 1/\epsilon
\end{equation}
Then $\Lambda$ represents the highest power of $z$ that will occur in
the growth law.

Let us make a change of variable.
We will replace $F(z)$ 
by the function $F((1+\epsilon)^{-1}z)$ and $G(z)$ by $G((1+\epsilon)^{-1}
z)$.  At the same time $f$ is replaced by
$(1+\epsilon) f((1+\epsilon)^{-1}z)$.  
Then the continuum law (\ref{Fceq}) for $F$ becomes
\begin{equation}
\label{Freg}
F_t=F_z \int \frac{d\theta}{2\pi} \Bigl(F_z((1+\epsilon) e^{i\theta}) {F}^{*}_z(
(1+\epsilon)e^{i\theta})\Bigr)^{-1} z \frac{z+e^{i\theta}}{z-e^{i\theta}}
\end{equation}
where now the cutoff dependence has been moved to the derivatives of $F$.
The continuum law (\ref{Gceq}) for $G$ becomes
\begin{equation}
\label{Greg}
\begin{array}{lll}
G_t&=&G_z \int \frac{d\theta}{2\pi} G((1+\epsilon) e^{i\theta}) 
G^{*}((1+\epsilon) e^{i\theta}) z (z+e^{i\theta})/
(z-e^{i\theta}) \\&& -G \int \frac{d\theta}{2\pi} G((1+\epsilon)e^{i\theta})
 G^{*}((1+\epsilon)e^{i\theta})
[z(z+e^{i\theta})/(z-e^{i\theta}]_z
\end{array}
\end{equation}
The value of $\epsilon$ must now be calculated.  

Before the averaging process, the dependence of the cutoff on $\lambda$
is easy to determine by, for example, expanding the function $f$ as
defined by equation (\ref{feq}).
After the averaging process, it is not necessarily the case that the
average cutoff $\epsilon$ will be determined in the same way from the
average value of $\lambda$.
The averaging may
introduce nontrivial behavior.  Instead I will look for the
dependence of $\epsilon$ on the cluster size; since the cluster size
and the average of $\lambda$ are related, this is an equivalent procedure.

Expanding the cutoff in the continuum
growth law to linear order in $\epsilon$  yields an additional
term in the equation for $F_t$.  This additional term changes equation
(\ref{norm}) to
\begin{equation}
\label{roceq}
{\rm Re}\Bigl(F_t \frac{|F_z|}{z F_z}\Bigr)
=1/|F_z| - 2 \epsilon {\rm Re}\Bigl(\frac{z F_{zz}}{F_z |F_z|}\Bigr)
\end{equation}

The additional term may be written as 
\begin{equation}
2\epsilon \, {\rm Im} \left(\frac {F_{\phi \phi}}{F_{\phi} |F_{\phi}|}\right)
\,=\,2\epsilon/R
\end{equation}
where $z=e^{i \phi}$ and $R$ is the local radius of curvature.
This is a surface tension term.

The basic idea will be to adjust $\epsilon$ to produce the correct
size for microscopic features; this size is the size of an
individual walker in the lattice formulation of DLA.

A dimensional analysis argument may help understand the size
of the cutoff.  This dimensional analysis argument will relate the
dependence of cutoff on macroscale to the dependence of cutoff on microscale.

The function $F(z)$ may be assigned the
dimension of length, and $z$ may be made dimensionless.  
This means that we are interpreting $z$ as a parametrization of the cluster.
Then both $\Lambda$ and
$\epsilon$ are dimensionless.
We know that 
$\Lambda$ must be a function of the size of an individual
walker, but then since
the dimensional argument implies that $\Lambda$ is
dimensionless,  $\Lambda$ must be proportional to some power of the ratio of
the size of the object to the size of an individual walker, as this
is the only way to form a dimensionless number.  Let $r_0$ denote the
length scale of an individual walker.

Let us see how to measure the linear size of the object.
Recalling the expansion of $F$ in Fourier coefficients given
by equation (\ref{FPS}),
by a theorem on univalent functions \cite{univalent}
the size of the object is at most 4 times $F_1$, the leading
term of the power series.  Asymptotically, $F_1$ will
measure the size of the object.  

To fix the minimum radius of curvature at $r_0$,
the size of an individual walker, the cutoff $\epsilon$ must be chosen so that
$\epsilon /R$, the surface tension term, 
 balances the electric field at the given radius.  To determine the
radius at which they balance, we need to make an assumption about the
singularities of $G$.  Let us assume $G$ has simple poles and therefore
$F_z$ has simple zeroes.  Suppose we look at points near a zero of $F_z$.
Without loss of generality, take this zero to be at point $z_0$
where $z_0=1-\delta$ with $\delta$ some small positive number.
Locally we find
\begin{equation}
F_z \propto z-z_0
\end{equation}
The electric field at $z=1$ is proportional to $1/\delta$.  The radius of
curvature at $z=1$ is proportional to $1/\delta^2$.  For the surface tension to
balance the electric field we require
\begin{equation}
\label{baleq}
1/\delta = \epsilon / \delta^2
\end{equation}
This implies that $\delta=\epsilon$.  Then, since $R=1/\delta^2$, we need
that $\epsilon=\sqrt{R}$.  If $R$ is equal to $r_0$, we find
\begin{equation}
\epsilon \propto r_0^{1/2}
\end{equation}

The dimensional argument then implies that
$\epsilon \propto (r_0/F_1)^{1/2}$, where now the proportionality
constant is dimensionless.
This implies that
\begin{equation}
\label{lf1}
\Lambda \propto (F_1/r_0)^{1/2}
\end{equation}
In the actual growth, $F_1$
is changing in time, but $r_0$ is constant.  Thus the time dependence
of $\Lambda$ is determined by $\Lambda \propto F_1^{1/2}$.  As expected,
the cutoff is increasing in time.
If the power series expansion for $G$ is defined by equation (\ref{GPS}),
then 
\begin{equation}
\label{yo}
G_0=1/F_1
\end{equation}
so that 
it is also possible to measure the
size of the object using the power series expansion of $G$.

As a further explanation of the dimensional analysis argument, it
may be directly shown that, if the cluster is approximately circular, with a 
small bump on it, then the dependence of $\Lambda$ on the cluster size
is correctly given by equation (\ref{lf1}).   The approximate circularity
means that instead of simply stating that the electric field is
proportional to $1/\delta$ and that the radius of curvature is
proportional to $1/\delta^2$, we keep track of the proportionality
constants in terms of $F_1$, and then directly show that $\epsilon\propto
(r_0/F_1)^{1/2}$.
The advantage of the
dimensional argument is that it is possible to make this argument
without any assumptions on the macroscale of the cluster; the dependence
of $\Lambda$ on $r_0$ follows from microscopic considerations, and the
dimensional analysis argument then yields the dependence on $F_1$.

One might worry that for the actual aggregate the poles will not
necessarily be simple poles.  As Halsey et. al. have shown \cite{Halsey1}, 
the surface is described by wedges with a non-zero opening angle,
and the singularities exist on a fractal set.  However, 
the continuum growth law of equation (\ref{Gceq}) only produces
simple poles in $G$.  It is only the
dynamics that lead apparently to the creation of non-simple poles, 
via an accumulation of simple poles.  Therefore, the
cutoff will be imposed as if the poles were simple.

Because the object grows, $\Lambda$ increases with time.  This
is what leads to nontrivial dynamics and to a fractal dimension less than 2.
If instead of varying with time, the cutoff $\Lambda$ were held constant,
 then the aggregate would asymptotically grow at a $\Lambda$-dependent rate 
proportional to the square root of time, and would have a fractal
dimension of 2. 
\subsection{Noise}
From now on, $G$ will be the function of interest and $F$ will be ignored.
There are two reasons for this: the continuum law for $G$ is simpler,
and noise may be more easily inserted into the law for $G$.
The actual growth of $G$ is not deterministic; we may write the
actual growth of $G$ symbolically as follows:
actual growth of $G$ = continuum growth of $G$ + (actual growth
of $G$ $-$ continuum growth of $G$).  The term in parenthesis represents noise.
This noise term will be written as 
\begin{equation}
s G^f(z) 
\end{equation}
where
$s$ is a constant with
dimensions of inverse time and $G^f$ is some function of $z$.
We approximate that $G^f$ vanishes on average.
If we expand $G^f$ in a series as $G^f_{-1}z^{-1} + G^f_{-2} z^{-2} + ...$,
we will write the average of $G^f_i G^{f*}_j$ as  
\begin{equation}
\label{Neq}
\langle G^f_i G^{f*}_j \rangle = \delta_{ij} N(j)
\end{equation}
where $N$ is some unspecified function.  We will also assume that any
average of several $G^f$ can be written as a product of pairwise averages.
These are the essential approximations in the noise.

With noise included, we modify equation (\ref{Greg}) to
\begin{equation}
\label{GN}
\begin{array}{lll}
G_t&=&G_z \int \frac{d\theta}{2\pi} G((1+\epsilon) e^{i\theta}) 
G^{*}((1+\epsilon) e^{i\theta}) z (z+e^{i\theta})/
(z-e^{i\theta}) \\&&- G \int \frac{d\theta}{2\pi} G((1+\epsilon)e^{i\theta})
 G^{*}((1+\epsilon)e^{i\theta})
[z(z+e^{i\theta})/(z-e^{i\theta}]_z + s G^f(z)
\end{array}
\end{equation}

The notation $G^f$ is used for the noise because, in the perturbation
theory, the function $G^f$ will play a role similar to a {\it free} field in
field theory.  

The approximations are justified for two reasons.  Since the noise
is essentially generated by the dynamics, that is a small amount of
noise will be amplified by the continuum growth, the dynamics should
not be very sensitive to how the noise is inserted.  This
means we need not worry about the exact form of $N(j)$.
Second, since the
actual growth law for $F^{-1}$ is rather simple, involving
a function acting on $F^{-1}$, it is most natural to
insert the noise into $F^{-1}$, or into $G$, which is the derivative of
$F^{-1}$.  Inserting the noise into the growth law for
$F$, which has a more complicated growth law,
may have a different effect on the overall dynamics.

\subsection{Continuum Limit In Momentum Space}
In order to make the RG calculations easier, I will also take a continuum
limit for the Fourier components of $G$ and $G^f$.  This will result in
replacing the discrete sums of equation (\ref{GPceq}) with integrals.
This amounts to a change in the geometry of the growth; instead of 
parametrizing the boundary of the growing cluster by a point on the
unit circle, we will parametrize it by a point on the real line in the
complex plane.  

In the neighborhood of a given point on the unit circle, such as
$z=1$, the unit circle is locally approximated by a straight line.  
As we look at shorter and shorter length scales, this approximation
becomes more and more accurate.  The equation $z=e^{i\theta}$ is 
approximated by $z=1+i\theta$.  Thus, on short scales, in the neighborhood
of $z=1$ we can approximate
\begin{equation}
\label{FFSE}
F(z)=F_1 z^1 +F_0 z^0 +F_{-1} z^{-1} +... \approx
 F_1 (1+i\theta) + F_0 + \int dj \, F(j) e^{-ij\theta}
\end{equation}
\begin{equation}
\label{GFSE}
G(z)=G_0 z^0 +G_{-1} z^{-1} +G_{-2} z^{-2} +... \approx
G_0 + \int dj \, G(j) e^{-ij\theta}
\end{equation}
where $F(j)$ and $G(j)$ should be considered 
as being defined by the above equations.  They are defined so that
$j$ is always a positive quantity.
This approximate form for $G(z)$ will break down for 
$\theta$ of the order of 1 radian.  This implies that the Fourier expansions
will break down for low values of $j$.  This has the effect of an
infrared cutoff; the cutoff will be at $j$ of order $\mu$, which is a
number of order 1.  The cutoff $\mu$ is constant in time, unlike $\Lambda$,
but under the RG we will find it convenient to rescale $\mu$.  After
introducing the equation of motion appropriate to this approximate
expression for $G$ I will then explain the effect of nonzero $\mu$ on
this equation.

{\it The symbol $j$ in equations (\ref{FFSE}),(\ref{GFSE}) 
will be referred to as
a momentum since it will play a role in the perturbation theory of the
next section equivalent to that of a momentum in a perturbation theory for
a field theory.} 

We take equation (\ref{GPceq}) and transcribe it to this continuum
approximation.  Noise is added as in equation (\ref{GN}).  The result is:
\begin{equation}
\label{ceq}
\begin{array}{lll}
G_t(j)& =& 
(1/\mu) \int dk \, t(j,k)\, G(k) \int \int dl \, dm \, G(l) G^{*}(m) \exp(-(l+m)/\Lambda) \\ &&\times \delta(k+l-m-j)
2 \theta(j-k) + s G^f(j)
\end{array}
\end{equation}
where  $t(j,k)$ is 
some general function (initially it is proportional to $(j-2k-1)$ as in
equation \ref{GPceq}), and where  
the factor of $1/\mu$ is inserted to produce the correct
dimensions for $t$ in the RG, as will be clear later.  
The insertion of the factor of $1/\mu$ simply
amounts to a redefinition of $t(j,k)$.
The function  $t$ will flow under the renormalization group.  
The exponential term is
the appropriate version of the ultraviolet cutoff in the continuum limit.
We define the continuous step function $\theta$ with by the same equation
(\ref{discretetheta}).

The effect of the cutoff $\mu$ is twofold: the $\delta$-function in
equation (\ref{ceq}) has a nonzero width of order $\mu$, and hence
a finite height.  The quantity $\delta(0)$ is of order $1/\mu$.
Also in the definition of the noise, equation (\ref {Neq}) is replaced by:
\begin{equation}
\label{Nceq}
\langle G^f(i) G^{f*}(j) \rangle = \delta(i-j) N(j)
\end{equation}
where again the $\delta$-function has a nonzero width.

By rescaling momentum, the cutoffs $\mu$ and $\Lambda$ may be
changed, but the ratio of the two
cutoffs will remain constant.  The purpose of the RG will be to integrate
the upper cutoff from $\Lambda$ to $\Lambda-\delta \Lambda$.  Then, for
the sake of convenience, the upper cutoff will be rescaled back to $\Lambda$.
The assumption is made that when $\Lambda$
is much greater than $\mu$ this renormalization does not change the essential
physics of the system.

Physically, equation (\ref{ceq}) describes the problem of DLA growth
in the upper half of the complex plane, where the boundary of the
growth is parametrized by a point $\theta$ on the real axis.  The cutoff
$\mu$ has the physical interpretation that growth only occurs in a
finite width on this axis.

For use later, let us define a functional $\widehat W$ such that
\begin{equation}
\label{Wdef}
\begin{array}{lll}
\widehat W \bigl[j, G(k), G(l), G^*(m)\bigr]& =&
(1/\mu) \int dk \, t(j,k)\, G(k) \int \int dl \, dm \, G(l) G^{*}(m) \exp(-(l+m)
/\Lambda) \\ &&\times \delta(k+l-m-j)
\end{array}
\end{equation}
Thus, the right hand side of equation (\ref{ceq}) is
$\widehat W \bigl[j,G(k),G(l),G^*(m)\bigr] + sG^f(j)$.  
The functional $\widehat W$ is linear in each of its last three
arguments.
\section{Adiabatic Assumption and Numerical Predictions}
\subsection{Adiabatic Assumption}
An important approximation is made, which changes the problem to
one of describing an aggregate which is statistically unchanging in time.
In the end, we will describe aggregates whose average size and roughness
remain constant.

The cutoff $\Lambda$ is slowly changing in time.  As the
object grows, $\Lambda$ changes more and more slowly.
The exact structure of the cluster at a given time depends upon 
its growth at all previous times, but since the object 
spends a long time growing with an approximately
fixed cutoff, it is expected that the structure of the object at a
given time $t$ with resulting cutoff $\Lambda$
is determined only by its growth during previous
times $t'$ with resulting cutoffs $\Lambda'$ such that $\Lambda'$ is
very close to $\Lambda$.  Times $t'$ such that $\Lambda'$ is very different
from $\Lambda$ will be so far in the past that we do not expect them
to alter the structure of the cluster.

Furthermore, if the cutoff is fixed, the equations of motion
are homogeneous, in the sense that up to a rescaling of time and noise,
two clusters, which differ only by a change of scale, will have
exactly the same growth for the same random noise.
More precisely, if $G(t)$ is a solution of the equation (\ref{GN}), with
fixed cutoff $\Lambda$ and given noise $G^f$, then, for any number $b$,
the function $bG(b^2 t)$ is a solution of equation (\ref{GN}) with the
same cutoff $\Lambda$ and with noise $bG^f$ and $s$ replaced by $b^2s$.

For a large object it is then reasonable to make
the  adiabatic assumption that, despite the changing cutoff, 
{\it up to a rescaling of the cluster,
the statistical properties of the function $G$ at 
some given time with some given cutoff $\Lambda$ are
well described by evolving an arbitrary initial $G$ for sufficiently long time
using the equations of motion with the cutoff 
held fixed at that value $\Lambda$}.

First, we will analyze the dynamics of $G$ in the fixed cutoff problem,
and then we will use the adiabatic assumption to relate it to the
changing cutoff problem.

Let us take equation (\ref{ceq}) and, holding the cutoff fixed, make
a dynamical rescaling of the function $G$ as it evolves under this equation.
After every time step of length $dt$ we will rescale $G$ to $G (1-s\, dt)$.
Then, the rescaled $G$ satisfies the equation of motion 
\begin{equation}
\label{rseq}
\begin{array}{lll}
s G(j) + G_t(j)&=& (1/\mu) \int dk \, t(j,k) G(k) \int \int dl \, dm \,
G(l) G^{*}(m) \exp(-(l+m)/\Lambda) \\&&\times \delta(k+l-m-j)
2 \theta(j-k) + s  G^f(j)
\end{array}
\end{equation}
On average, the rescaled $G$ has constant size.

The amount by which the cluster is rescaled per unit time, $s$, is
the same $s$ referred to in the section on Noise.  It is simply
a matter of notational convenience to chose these two numbers $s$ to be
the same.  Any other choice of $s$ in the section on noise simply amounts
to a redefinition of $G^f$. 

The function $G$ before rescaling is growing in time.  To determine how
rapidly an unrescaled
cluster of given size and given, fixed  cutoff grows,
we may follow this procedure: evolve a rescaled
cluster using equation (\ref{rseq})
with an $s$ chosen such that the rescaled cluster is of the desired size.
Then, from the value of $s$ needed to maintain the desired size, 
determine the growth rate of the unrescaled cluster.  For the
unrescaled cluster, the average
of $\frac{d\log(G_0(t))}{dt}=s$.

Under the adiabatic assumption, we can now use the growth rate for the
fixed cutoff problem, this growth rate being 
a function of the size of the cluster
and the cutoff, and use it to determine the growth rate for a cluster
with a changing cutoff. 
To determine the growth rate of a cluster of given size
in the changing cutoff problem 
one can determine the cutoff from the size of the cluster and then
calculate the growth rate of a cluster of the same size in the
fixed cutoff problem, using the rescaling trick to determine
how quickly that cluster grows.

Under RG scaling, in fact, we will find that the equation of
motion changes in such a way that $s$ changes; in fact, $s$ may
acquire momentum dependence.
Before RG scaling $s$ will be negative,
since $F$ is increasing, causing $G$ to decrease.  After RG flow, an
appropriate combination of $t,s,$ and $G^f$ goes to a universal value.
The adiabatic assumption
will mean that we assume that at every instant in the original DLA problem,
the function $G$ is described by a function in the long time limit of
the problem with a fixed cutoff and a rescaling term $s$, where $s$ is
picked to obtain the correct overall scale for $G$.  

On average, $G_t$ in equation (\ref{rseq}) vanishes.  What is
left of $G_t$ after the rescaling process is just fluctuations
about the average growth.  There may in
fact be solutions such that $G_t$ vanishes identically, but this is
unimportant.
In the RG, even with $G_t$ non-zero, the 
$s$ term and interaction term ($G^3$ term)
will determine the nature of the aggregate.  As will be shown, under RG flow,
the $G_t$ term flows under RG so that, the fluctuations in $G_t(j)$ decrease
as $j$ decreases.
The lower momentum terms then, in the unrescaled
problem, will have their growth more accurately given just by the $sG$ term.
Fluctuations about this overall growth will be less.
\subsection{Numerical Predictions}
It will be worthwhile to mention at this point that already some
definite numerical predictions can be extracted from the above work.
Since the RG that follows relies upon the continuum equations, it is good
to independently check the validity of these equations for describing
DLA.  

If any RG is to hold, the coefficients of $G$ must obey some scaling law.
In the last section of this paper, such a scaling law will be shown 
numerically.  The coefficients decay with a power law.
It will be the purpose of the RG to calculate this power
law.

Since the absolute value of $G$ is equal to the local electric field,
there is a close connection between $G$ and the multifractal exponents
of references \cite{mulf,mulf2,mulf3}.  These exponents are defined by
equations (\ref{tqeq}),(\ref{dqd}).
The $(2n+1)$-th power of the electric field, integrated
over the object, is given by $\int d\theta (G(\theta) G^*(\theta))^n$.  
This integral over $\theta$ can be converted to an integral over
components of $G$ in momentum space.  Cutting those off at 
momenta $\Lambda$ is equivalent to cutting the real space integration
off at a length scale of order $\Lambda^{-2}$.  Since the
multifractal exponents are defined in terms of the scaling of powers
of the electric field against length, we obtain an equivalent definition
of multifractal exponents in terms of scaling of powers of $G$ against
cutoff.  This discussion of exponents will be done in more detail later,
after the RG permits us to calculate these exponents analytically.

As a check of the adiabatic assumption, a numerical simulation
was performed using the original discrete conformal mapping model defined
in section (1.2).  In this
simulation, after every growth step, the object was shrunk by some constant
factor.  Visually we could not see any difference, in the growing
region on the surface, between the cluster
shrunk after every growth step and another cluster which was not shrunk.
A calculation of fractal dimension also failed to show any significant
differences.
\section{Perturbation Theory}
A perturbation theory is developed for the equation of motion (\ref{rseq}).  
This permits
in principle the calculation of any correlation function of the theory in
terms of noise averages.  In practice, a resummation of the series is employed
which expresses multi-point correlation functions in terms of two-point
correlation functions.

\subsection{Perturbation Rules}
Using the adiabatic assumption, the equation of motion (\ref{rseq})
looks very much like the variation of
an action.  Although there no such action can be found, 
a perturbation theory will be developed, based on this analogy, to permit
the use of techniques from field theory.

This technique is very similar to that used, for example,
in solving the Navier-Stokes equation \cite{NSP}.  Such a perturbation
theory has been known for many years.  Before proceeding with the details,
let me summarize the essential attributes: a perturbation theory
is developed by expanding $G$ in powers of the noise, $G^f$, and
expanding correlation functions of $G$ in terms of two-point 
correlation functions of the noise.
Since the noise is amplified by the dynamics of equation (\ref{rseq}), this 
expansion is not expected to converge.  However, after resummation of the
series, it becomes possible to replace this by an expansion not
in the two-point correlation function of the noise, but in the two
point correlation function of $G$ itself.  Further, a resummation of
series leads to a resummed propagator (defined below). Unlike the
Navier-Stokes perturbation theory for turbulence\cite{NSP}, it  will not not be
necessary to define a resummed interaction vertex.
The above procedure leads to a well defined perturbation series in 
``skeleton" diagrams.  One point that will be necessary in the
following treatment that is not necessary in the case of turbulence
is that $G$ is a complex field, and thus the propagators will be
directed lines; the notation that follows will therefore differ
from that seen in Navier-Stokes problems.

The perturbation theory is constructed as follows: the equation of motion 
(\ref {rseq})
permits us to solve for $G(j)$ in terms of a cubic in $G(j)$.  
Using the definition of $\widehat W$ in equation (\ref{Wdef}) we write
\begin{equation}
\label{p1}
G(j)=\Bigl(s(j)+\frac{\partial}{\partial t}\Bigr)^{-1}\left(\widehat
W\bigl[j,G(k),G(l),G^*(m)\bigr]+s(j)G^f(j))\right)
\end{equation}
where $s(j)$ is used instead of $s$ because $s$ may, under RG flow,
acquire $j$ dependence.

The operator $(s(j)+\frac{\partial}{\partial t})^{-1}$ may be
expanded as a formal power series in $\frac{\partial}{\partial t}$.  
In Navier-Stokes perturbation theory, this operator is
referred to as the propagator, in analogy with a similar object in field theory.

We may then iteratively solve equation (\ref{p1}) as follows
\begin{equation}
\label{p2}
\begin{array}{lll}
G(j)&=&\frac{1}{s(j)}\widehat
W\bigl[j,G(k),G(l),G^*(m)\bigr]+G^f(j)+{\rm O}(\frac{\partial}{\partial t})\\
&=&\frac{1}{s(j)}\widehat W\bigl[j,G^f(k),G^f(l),G^{f*}(m)\bigr]+G^f(j)\\
&& + \frac{1}{s(j)}\widehat W\bigl[j,\frac{1}{s(k)}\widehat W[k,G(n),G(o),G^*(p)],G^f(l),G^{f*}(m)]\\
&& + ...
\\&&+ {\rm O}(\frac{\partial}{\partial t})\\
&=&...
\end{array}
\end{equation}
The iterative solution of equation (\ref{p1}), by solving for the values
of $G(k)$, $G(l)$, and $G^*(m)$ on the right hand side, is
an expansion in powers of the interaction, $t$.  This is simultaneously
an expansion in powers of the noise, $G^f$.  The zeroth order and first order
terms in $t$, and
one of the second order terms, have been written in equation (\ref{p2}).
At any
point in the process, one can stop the perturbation expansion by setting
$G(j)=G^f(j)$ plus higher orders in $\frac{\partial}{\partial t}$.  Thus, in
the perturbation expansion for $G$ one sums at every stage over two possible
expressions for $G$:
\begin{equation}
\begin{array}{lll}
G(j)&=& (1/s(j)) (1/\mu) \int dk \, t(j,k) G(k) \int \int dl \, dm \,
G(l) G^{*}(m) \exp(-(l+m)/\Lambda) \\ &&\times \delta(k+l-m-j)
2 \theta(j-k) + {\rm O}(\frac{\partial}{\partial t})
\end{array}
\end{equation}
or
\begin{equation}
G(j)=G^f(j) + {\rm O}(\frac{\partial}{\partial t})
\end{equation}

These operations can be
represented graphically with Feynman diagrams, in which
$t$ is an interaction term and
$s$ is like a mass term.   This leads to a series of diagrams for $G$.
These diagrams have no loops.

The quantities computed as described above depend upon the specific realization
of the noise.  Since we are interested in average quantities,
we will compute correlation functions.
A correlation function is defined as
an average over noise of a product of several
$G(j),G^*(j)$ with the same total number of $G$ and $G^*$,
and will be written as
\begin{equation}
\label{cordef}
\langle G(j_1) G(j_2) ... G(j_n) G^*(k_1) G^*(k_2) ... G^*(k_n) \rangle
\end{equation}
When computing 
these averages, the average over noise must be taken.  This is done
by taking noise terms resulting from the above expansion for $G$ and $G^*$
and contracting them with each other in all
possible pairwise fashions.  Each contraction of two noise terms at
momenta $j,k$
leads
to a factor of $\delta(j-k) N(j)$, as given by
equation (\ref{Nceq}).  This leads to diagrams with loops.

Fig.~3 indicates how diagrams for the theory are drawn.  
There are three types of diagrams that may be drawn.  They are drawn
in essentially the same fashion, except that different meanings are
assigned to the external lines and different numbers of noise contractions
are included.  A diagram for a correlation function has
one external line
for each term in the correlation function to be computed.  Diagrams for
quantities other than correlation functions may have different
meanings assigned to the external lines.  These diagrams all involve one or more
external lines being expressed as a function of the other external lines.
This may occur either as an expansion of $G$ directly in terms of the noise,
or as a piece of a diagram that occurs inside another diagram.
For example, the diagram (a)
in Fig.~3 is not itself a correlation function, but it represents a
term that may occur inside a computation of a correlation function.  
For this diagram, the line for $G(j)$ is expressed in terms of other lines,
which in turn may be set equal to the noise, or may be further expanded.

The notation for diagrams is the following: crosses denote
contractions on noise, and the directions
of the lines indicate complex conjugation
and orientation within the diagram.  A line may be said to carry
momentum $j$; when two lines are contracted, they must carry the
same momentum and all external lines are assigned a momentum determined
by the particular correlation function to be computed.
A $t$ vertex has 2 lines entering and 2 lines leaving, corresponding
to $G$ and $G^*$.  The $G(j)$ line
for a $t(j,k)$ vertex is drawn as entering, the $G^*(m)$ is drawn as entering
since it is complex conjugated, while the other lines are drawn as leaving the
vertex.  The $j$ and $k$ lines will always be drawn parallel to each other
in a $t$ vertex; the $l$ and $m$ lines will be drawn at an angle.
The distinction must be made between $j,k$ and $l,m$ because $j$ and $k$
enter into $t$ while $l$ and $m$ are summed over blindly.

The rules for diagrams for correlation functions may be summarized as follows:
to calculate the average of the product of
$G(j_1) G(j_2) ... G(j_n) G^*(k_1) G^*(k_2) ... G^*(k_n)$, where
$j_i,k_i$ are various numbers: draw one external
line for each term in the correlation function.  The terms in $G$ should be
drawn as entering the diagram with momentum $j_i$, 
while those for $G^*$ should be drawn
as leaving the diagram with momentum $k_i$. 
Draw all possible diagrams, assigning a factor
of $t(j,k) 2 \theta(j-k)$ to each vertex and a factor of $1/s(j)+
{\rm O}(\frac{\partial}{\partial t})$ for each line, 
while conserving the sum of ingoing and outgoing
momentum at each vertex.  For each noise contraction assign a factor
of $s^2(j) N(j)$.   Finally, the perturbation theory may give rise to
a term such as $\theta(0)$.  This will arise from
something like $\int dj \delta(j-k) \theta(j-k)$.
While $\theta(j)$ is 0 for negative $j$ and
1 for positive $j$, $\theta(0)$ will be taken to be equal to $1/2$,
according to equation (\ref{discretetheta}).  This
is a result of the factor of two difference between powers of $z$ in the
expansion for $f$; this difference was discussed in
reference to equation (\ref{fsle}).
Alternately, if we recall that all $\delta$ functions should
be assigned some nonzero width, then the above integral evaluates to $1/2$.
\subsection{Resummation}
This theory exhibits a spontaneous breaking of
circular symmetry.  One starts the growth process with a circularly
symmetric cluster, which implies that $G(j)=0$ for $j>0$, but this is not a 
stable state.  Instead, the dynamics evolves $G$ to one of an
infinite number of states with nonzero $G(j)$, although on
average $G(j)=0$ for $j>0$.
If we impose some boundary condition,
such as $G=1$ at time $t=0$, and look at $G$ for much later times,
the perturbation theory in a small noise term causes us to
to reach large values of $G$.
Having imposed these
boundary conditions, the noise is amplified by the dynamics and will grow
large.

This large growth of noise means that if we evaluate the two-point
correlation function $\langle G(j) G^*(k) \rangle$ we will obtain some answer 
of the
form $\delta(j-k) |G^2(j)|_{av}$, where $|G^2(j)|_{av}$ is an appropriate function of $j$,
and $|G^2(j)|_{av}$ is much larger than $N(j)$.  In the long time limit,
$|G^2(j)|_{av}$ is not a function of time.

Then, we may imagine that, when calculating any other
correlation function, at any stage
in the perturbation theory, two lines which were contracted to obtain
the value $N(j)$, can instead have their contraction dressed with
additional interactions to convert $N(j)$ to $|G^2(j)|_{av}$. 
Two important points must be made about this procedure.
First, it is important that this value $|G^2(j)|_{av}$ is completely
uncorrelated with any other values of $G$ in the diagram; 
all correlations are already taken into account by the interaction vertices
of the diagram.  Second, we must be careful to avoid overcounting; since
each contraction in a diagram includes many diagrams involving dressing
the contraction in various ways, we must not further dress these
contractions.

For notational convenience, I will continue to
write $G^f$ everywhere, but now 
\begin{equation}
\label{resum}
\langle G^f(j) G^{f*}(k)\rangle =\delta(j-k) 
|G^2(j)|_{av}
\end{equation}
  $G^f$ is similar to a free field in field theory.  Any
average product of several $G^f$ can be decomposed into pairwise products.
This is the resummation of noise into two-point correlation functions
of $G$ that was mentioned above in the comparison to
Navier-Stokes perturbation theory.  $G^f(j)$ is now an unknown function
of $j$; one of the goals of the RG will be to determine the
dependence of this function on $j$.

Similarly, the function $s(j)$ can be resummed.
Any single line between two interaction vertices, which 
would normally be represented by a factor of $1/s$, will instead be
represented by $1/s^{eff}$, where $s^{eff}$ takes into account possibilities
of dressing that line, without interactions with other lines.   $s^{eff}$ is
an ``effective" $s$.
This is the resummation of propagators.

As a result of these two resummations, to avoid overcounting
we must require that we do not count diagrams in which some portion of
the diagram that contains interactions 
has only two lines leaving it.  See Fig.~4 for examples of
contributions to $j$,
contributions to $s^{eff}$, and diagrams which can not be included in the
theory as they would overcount contributions.

As one will be able to verify after performing the RG
calculation, within the formal power series expansion
of $(s+\frac{\partial}{\partial t})^{-1}$, one may
neglect all terms in $\frac{\partial}{\partial t}$ in the
RG calculation of the next section.  Terms only flow
to higher powers of $\frac{\partial}{\partial t}$.  Because $|G^2(j)|_{av}$
is taken to be constant in
time, terms with $\frac{\partial}{\partial t}$ will drop out in many places.

Then, the perturbation rules, ignoring time derivative, may be summarized
 as follows: draw all diagrams (subject to the rules forbidding
overcounting), with appropriate external lines, assigning
a factor of $1/s^{eff}(j)$ to each line, and a factor of $t(j,k) \theta(j-k)$
to each vertex, and a factor of $(s^{eff})^2(j) |G^2(j)|_{av}$ to each contraction on
noise, while conserving momentum everywhere.

\subsection{Example Calculation}
It will be useful to give a simple example of applying such a
perturbation theory to a non-interacting system.  For example, consider
the following simplified equation of motion:
\begin{equation}
G_t(j) + s G(j)=t(j,j) G(j) G(0) G^*(0) + s G^f(j)
\end{equation}
where $t(j,j)$ is proportional to $j-2j-1$ which is equal to $-j-1$.
Let us suppose $G^f$ is constant in time, to simplify the problem
further.
Then, let us define the time scale so that $t(j,j)=-j-1$.  
Let us pick the desired scale of the cluster so that $G(0)=1$.
Physically, $G(0)$ is very large compared
to the noise $G^f(0)$.  This mean that for a stationary state (after all,
$s$ is adjusted to produce a stationary average size) we may
let  $G^f(0)$ be small and we need  $s$ approximately equal to $-1$.

Now that the values of $s,t$ are
fixed, we may find the solution, either using perturbation theory
or using a straightforward solution.  The latter method
gives $-G(j)=-G(j)-j G(j) -G^f(j)$.  Therefore, $G(j)=G^f(j)/j$
and $\langle G(j) G^*(j)\rangle =N(j)/j^2$.  This noninteracting system is 
stable.

The perturbation theory for $G(0)$ is slightly tricky.  If we ignore
the time derivative, we can only reach small values of $G(0)$ in
the perturbation expansion.  However,
in fact, for the given $s$ and $t$, a small $G^f(0)$ produces a large $G(0)$,
under the time evolution.  Let us suppose that this part of the process
has been done, and that we obtain a resummed expression for 
$\langle G(0) G^*(0)\rangle$.
This resummed expression is $|G^2(0)|_{av}=1$.
Then, we may ignore the time derivatives and obtain the expression
for the higher $G(j)$ in a straightforward fashion.

The perturbation theory expansion to $G(j)$
gives the following infinite sum:
\begin{equation}
\label{exam}
G(j)=G^f(j)+(|G^2(0)|_{av}t(j)/s) G^f(j) + (|G^2(0)|_{av}t(j)/s)^2 G^f(j) + . . . 
\end{equation}
This is equal to $G^f(j) (1-|G^2(0)|_{av} t(j)/s)^{-1}=G^f(j)/j$.  This perturbation
expansion is shown in Fig.~5.  Each line terminates by setting $G$
equal to the noise.
Similarly, the perturbation expansion to the correlators gives a product
of two infinite sums.  This product is represented in diagrams by taking
the sum in Fig.~5 and noise contracting it with its complex conjugate.

One may define $s^eff(j)$ for this simple theory; the
diagrams of Fig.~5 define the inverse of $s^{eff}$.
Therefore
\begin{equation}
s^{eff}(j)=s-t(j)|G^2(0)|_{av}
\end{equation}
which simplifies equation (\ref{exam}) to
$G(j)=\frac{1}{s^{eff}(j)}sG^f(j)$.
\section{Renormalization Group Calculation}
We investigate the effect of changing the cutoff in the equation of motion.
This leads to the  introduction of new diagrams to describe the changes
in the theory as a result of lowering cutoff.  It is shown how to
incorporate these into a change in $s,t$.  The fixed point is found. 

\subsection{Lowest Order Contributions to RG Flow} 

In the RG calculation, first the cutoff is lowered from $\Lambda$ to 
$\Lambda-\delta \Lambda$.
 If the cutoff is
imposed in a smooth fashion (interaction term $t(j,k) G(k) G(l) G^*(m)
\exp[-(l+m)/\Lambda]$), the
 change in the theory under a change in the cutoff can be obtained
by adding an additional term to the equation of motion (\ref{rseq}), 
equivalent to
the original cubic term, except that the cutoff
$\exp[-(l+m)/\Lambda]$ is replaced by $(\delta\Lambda) \partial (\exp[-(l+m)/\Lambda])/\partial\Lambda$.   This new term is
\begin{equation}
\label{rgeq}
\begin{array}{l}
(1/\mu) \int dk \, t(j,k)\, G(k) \int \int 
dl \, dm \, G(l) G^{*}(m) (\delta\Lambda/\Lambda)
(-(l+m)/\Lambda) \exp(-(l+m) /\Lambda)
 \\ \times \delta(k+l-m-j)
2 \theta(j-k) 
\end{array}
\end{equation}
The sum of the original interaction term plus
this new term is equal to the interaction term at reduced $\Lambda$.
This term will have a circle around the vertex
when it appears in a diagram.  

Because this term is small when the external momenta $l,m$ are
small, it does not directly enter into correlation functions of the
low momentum theory.  It enters indirectly, in more complicated diagrams.
We will then consider various such diagrams
which include this term, and show that, for the simplest such diagrams,
these diagrams may be rewritten in terms of a change in $s$ and $t$.

This procedure has some slight logical differences with other
RG procedures.  In other procedures, the cutoff is often imposed in
a sharp fashion at some momentum.  Here, the cutoff is imposed in
a very smooth fashion; I believe this has certain logical advantages.
This procedure is similar to the technique of counterterms in the
original formulation of renormalization in field theory.

First, I will evaluate the simplest diagrams to which this new term gives
rise, and do the RG calculation, then in the next
subsection I will consider other possible
diagrams and explain why they are neglected.  Throughout the
RG calculation, when I write $|G^2(\Lambda)|_{av}$, I
actually mean
\begin{equation}
\int dl \, |G^2(l)|_{av} \frac{\partial e^{-2l/\Lambda}}
{\partial\Lambda}
\end{equation}
This is just a weighted sum of $|G^2(l)|_{av}$ at $l$ of order $\Lambda$, and
due to the smooth cutoff it is this weighted sum that will enter into
all the diagrams considered.

The diagrammatic expansion for $s^{eff}$ will now include the diagram
shown in Fig.~6.
This   changes $s^{eff}(j)$ to 
\begin{equation}
\label{srg}
s^{eff}(j)-t(j,j)|G^2(\Lambda)|_{av}\delta\Lambda/\mu
\end{equation}
The term $2 \theta(j-k)$ in equation (\ref{rgeq})
gives 1 in this case, as discussed in the perturbation theory rules.

Additionally, the new term in the equation of motion can give rise to
a diagram as shown in Fig.~7a, which can be represented by changing
 $t(j,k)$.
This arises from substituting 
\begin{equation}
\label{subs}
s^{eff}(l) G(l) = G^f(m) t(l,m) G(n) G^*(o) 
\delta (m+n-l-o) 2 \theta(l-m) \end{equation}
and 
\begin{equation}
G^*(m)=G^{f*}(m)
\end{equation}
 into equation
(\ref{rgeq}), taking $l$ of the order of $\Lambda$.  The 
result changes $t(j,k)$ to 
\begin{equation}
\label{trg}
t(j,k)+2 (\delta \Lambda/\mu) 
t(j,k) t(\Lambda,\Lambda+k-j) |G^2(\Lambda+k-j)|_{av}/s^{eff}
\end{equation}

Assuming $\Lambda$ is
high momentum and $j,k$ are low momentum, then $\Lambda+k-j=m$ is high
momentum.  One does not use terms arising from substituting 
$s^{eff}(l) G(l) = G(n) t(l,n)
G^f(m) G^*(o)$, instead of the
substitution of equation (\ref{subs}),
because such terms involve too many high momentum components
of $G$.   The diagram corresponding to this term is shown in Fig.~7b.
Such terms should be ignored, as they will be small
when determining the behavior of the system for momenta
much less than $\Lambda$.  Remember that the
cutoff is imposed, in the original equation, on $G(l)$ and $G^*(m)$, not $G(k)$.
This means that the presence of the high momentum term $G^*(m)$ will
make such terms small.

All momenta are now rescaled by
$\frac{\Lambda}{\Lambda-\delta \Lambda}$
to put the ultraviolet cutoff back at $\Lambda$.  This
changes $s^{eff}(j)$ to 
\begin{equation}
\label{srs}
s^{eff}(j)-j(d s^{eff}(j)/d j)
(\delta\Lambda/\Lambda) 
\end{equation}

Because of the integration in the interaction term, and the one power of 
$\mu$ extracted from $t(j,k)$, the dimension of
$t(j,k)$ is equal to $({\rm momentum})^{-1}$.
Therefore, under rescaling, the function $t(j,k)$ becomes 
\begin{equation}
\label{trs}
t(j,k)-j(\partial t(j,k)/\partial j)(\delta\Lambda/\Lambda)-k(\partial t(j,k)/\partial k)(\delta\Lambda/\Lambda)-t(j,k)(\delta \Lambda/\Lambda)
\end{equation}

Combining the
terms resulting from the integration, equation (\ref{trg}),
with those from the momentum rescaling, equation (\ref{trs}),
every term in the change of $t(j,k)$ either is a function of $(j-k)$, or
would be a function of $(j-k)$ if $t(j,k)$ were a function of $(j-k)$.  For a
stationary point, then require that 
\begin{equation}
t(j,k)=t(j-k)
\end{equation}
If $t(j,k)=t(j-k)$, then
similar logic using equations (\ref{srg}),(\ref{srs})
requires that $s^{eff}(j)$ becomes a constant.
These requirements of momentum independence of $s,t$ 
likely do not hold at the extreme infrared for any true system, or
else the equation of motion (equation (\ref{rseq}) using renormalized
$t$ and $s$) would have no non-zero solution, but in the
scaling region between the infrared and ultraviolet, they will hold.

For the rest of this subsection, the number $s$ is the constant to 
which $s^{eff}$ flows under the RG, and the number $t$ is $t(0)$.

At this point, I have done the first part of the RG for the two numbers $s$ and $t$. It is now necessary to rescale $G$ to leave equation (\ref{rseq})
invariant under the renormalization and rescaling.
Let $G$ be rescaled to  
\begin{equation}
\label{GRS}
(1-r(\delta \Lambda/\Lambda))G
\end{equation}

This implies a rescaling of $s$ to
\begin{equation}
\label{sgrs}
(1+r(\delta \Lambda/\Lambda))s
\end{equation}
and a rescaling of $t$ to
\begin{equation}
\label{tgrs}
(1+3r(\delta \Lambda/\Lambda))t
\end{equation}

Naively one might expect that there could also be an overall rescaling
of both $s$ and $t$ by a factor $a$. This would leave $G$ unchanged.  
Taking derivatives of $\log(s)$ and $\log(t)$ with respect
to $\log(\mu)$ (since the ultraviolet cutoff
is lowered to $\Lambda-\delta \Lambda$ and then rescaled back to $\Lambda$,
it is actually $\mu$ that
changes in this process), for $s$ and $t$ to be stationary we find:
\begin{equation}
\label{fixed}
r-(\Lambda/\mu)t|G^2(\Lambda)|_{av}/s+a=0
\end{equation}
\begin{equation}
\label{fixed2}
3r+2(\Lambda/\mu)t|G^2(\Lambda)|_{av}/s+a-1=0
\end{equation}
This in itself does not provide enough information to extract anything useful,
because having both $r$ and $a$  means that $s$ and $t$ can be scaled to
arbitrary values.  

However, in fact, $a=0$, as will now be shown.
The time derivatives of equation (\ref{rseq}) were ignored in the RG because
they did not effect the flow of $s$ and $t$.  We can still use arguments
about the time derivative, $G_t(j)$, to show that $a=0$.
We can use equation (\ref{rseq}) to calculate $G_t(j)$ of the
renormalized, rescaled problem in two ways.
One way is to to take $G_t(j)$ for the problem before renormalization and
rescaling, and then rescale the momentum $j$ and multiply $G_t(j)$ by 
$(1-r(\delta \Lambda/\Lambda))$ in analogy with equation (\ref{GRS}).
Another way is to use the renormalized, rescaled $s$, $t$, and $G^f$ 
in equation (\ref{rseq}) to compute $G_t(j)$.  
For these two methods of
computing $G_t(j)$ to give the same result, as required by the fact that they
describe the same system, 
it may be shown that
\begin{equation}
\label{azero}
a=0
\end{equation}
A simple way of stating the argument leading to this result is that, although
if time derivatives are neglected in equation (\ref{rseq}) the equation
is invariant under multiplying 
the three terms $s$, $t$, and $G^f$ (here referring to the noise
before resummation, that is, the noise described by equation \ref{Nceq})
 by the same constant, such a multiplication
does not leave the time derivatives invariant and so should not be allowed in
the rescaling part of the RG.

Then, the fixed point equations (\ref{fixed}),(\ref{fixed2})
can be solved for 
\begin{equation}
\label{dimcoup}
r=(\Lambda/\mu)t|G^2(\Lambda)|_{av}/s=1/5
\end{equation}
This result for $r$ is the main result of the first order RG.

This gives the rescaling of $G(j)$ with $j$.  The magnitude of
$G(j)$ must decay as $j^{-1/5}$.  Thus
\begin{equation}
|G^2(j)|_{av}\propto j^{-2/5}
\end{equation}
The combination $(\Lambda/\mu)t |G^2(\Lambda)|_{av}/s$ is invariant
under a rescaling of $G$ of $G$ by equation (\ref{GRS})
and a corresponding rescaling of $s$ and $t$ by
equations (\ref{sgrs}),(\ref{tgrs}),
and gives us the dimensionless coupling
constant for this problem.  From equation (\ref{dimcoup}) the dimensionless
coupling constant is $1/5$, which is not infinitesimal;  however, after RG flow,
the problem is no longer strongly coupled, as the constant is less than 1.

The reason  that
a factor of $1/\mu$  was removed
from $t(j,k)$ is now clear; this makes the above coupling
constant truly dimensionless.  As a result of the removal of the
factor of $1/\mu$, the dimension of $t(j,k)$ is $({\rm momentum})^{-1}$.
However, $|G^2(j)|_{av}$ has the dimension of momentum; this is
because, taking $G$ to be dimensionless, equation (\ref{resum})
gives $|G^2(j)|_{av}$ a dimension inversely proportional to the 
$\delta$-function.  The $\delta$-function has dimension of inverse
momentum, and thus the end result is to make $t|G^2(j)|_{av}$ dimensionless.

As a further comment on the dimensionality of the coupling constant,
recall that the $\delta$-functions have a finite height proportional to $1/\mu$.
This finite height changes under the RG, which implies a rescaling of
$|G^2(j)|_{av}$ under the RG flow; by multiplying $t$ by $1/\mu$ we
shift this rescaling onto $t$.

An argument was made above, leading to equation (\ref{azero}),
involving $G_t$ for the problem before and after RG.
We may
extend this argument and
also say something about the magnitude of $G_t(j)$ for different $j$.
The quantitities $sG$ and $tG^3$ remained constant under the RG, as a  
result of the rescalings of $G$, $s$, and $t$ 
and the renormalizations of $s$ and $t$, from equations
(\ref{srg}),(\ref{trg}),(\ref{srs}),(\ref{trs}),(\ref{GRS}).
We also need $G_t$ to remain constant since this is also a 
term in equation (\ref{rseq}).  Suppose the characteristic inverse
time scale for fluctuations in $G(j)$ is $\omega(j)$.  Then $G_t(j)$ is of
order $\omega(j) G(j)$.  For this combination to remain constant, 
$\omega$ must change as a result of the rescaling of $j$ in the RG.
In fact,  $\omega(j)$ must have the same log derivative
under RG that $s$ does, although the log derivative of $s$ results from
renormalization (equation \ref{srg})
while the log derivative of $\omega(j)$ results from
rescaling.
This implies that
\begin{equation}
\label{rgw}
\omega(j)\propto j^{1/5}
\end{equation}
This means that for smaller $j$, the time scale for fluctuations is
longer.  Returning to
the original problem, as described by equation (\ref{GN}) with a time-dependent
cutoff, 
this means that the lower Fourier coefficients grow at a roughly constant rate.
This self-consistently justifies the adiabatic assumption of section 3.
\subsection{Other Contributions to RG Flow}
One may imagine that the new term of equation (\ref{rgeq}),
representing the rescaling of the cutoff, may enter into additional
diagrammatic contributions.  Various possibilities are shown in Fig.~8.
I will show that, for low momentum behavior, these terms are unimportant and
then discuss in more generality why other contributions are negligible.

One may check by hand that the first example is small if external
momenta are much less than $\Lambda$.  The second
example will be discussed below.  The third example
vanishes due to phase space factors.  The fourth example vanishes due to
phase space factors if the two lines leaving the top of the diagram are
close in momentum; this means it does not alter the RG flow of $t(j,k)$
when $j=k$.  In the rest of the section, various other diagrams will
also be said to ``vanish"; this will only mean that they vanish when
considered either at low momentum or, if they contribute to the RG flow
of $t(j,k)$, when considered at $j=k$.

The fifth example, a contribution
to the six-point function, will be seen below 
to be small when calculating
correlations of only 4, and not 6 or more, $G(j)$.  The sixth
example should not be considered when the rescaling of $|G^2(j)|_{av}$ is
taken into account; since $(\Lambda/\mu) |G^2(\Lambda)|_{av}$ is stationary under RG flow, such
a diagram is canceled by the various rescalings.

In order to indicate in general why such contributions may be neglected,
I would like to define some additional terminology to describe certain
paths and contractions in these diagrams.
When considering a contribution to $s^{eff}$, one may follow one line through
the diagram as follows: start with  the incoming line.  At every $t$ vertex,
if one enters with $G(j)$, follow out along $G(k)$, not $G(l)$ or $G^*(m)$,
where the roles of $j,k,l,m$ are as in equation (\ref{rseq}).
If the line one is following is contracted (this will be referred to as 
an {\it exceptional contraction}) 
with a $G^*(m)$ leaving a $t$ vertex, follow out
along $G(l)$ of that $t$ vertex.   
This path will be referred to as the {\it main line}.
Now, any diagram that includes an exceptional contraction, such
as the diagram of Fig.~8a, will
be small for low momenta, since the $G(k)$ leaving such
a diagram will only have a small range over which to integrate.
For Fig.~8a the main line is given by following the horizontal arrow
along the bottom of the diagram from left to right, through the $t$ vertex,
until it bends up and left.  Then go down and left through the noise contraction
into the $t$ vertex, and then leave the $t$ vertex along the line going
up and left.  Follow this line through its bend back to the right until
it leaves the diagram.

The smallness, of the contribution to $s^{eff}$ given in Fig.~8a,
for small momenta does not completely justify
the neglect of such terms.  For example, when evaluating the RG 
contribution to $t$, the value of $s^{eff}$
used is $s^{eff}(\Lambda)$, not the low momentum $s^{eff}$, and thus
a high momentum contribution to $s^{eff}$ may change the low
momentum renormalized $t$.  However, even for a calculation of
$s^{eff}(\Lambda)$, the exceptional contraction will mean some reduction in
available phase space over which to integrate.

One may check that contributions to $s^{eff}(j)$ 
like the third example in Fig.~8
will always vanish, regardless of what $j$ is,
due to the $\theta$ functions in equations (\ref{rseq}),(\ref{rgeq}).  
The lines coming off
of $t$ vertices connected to the main line must be contracted within
themselves, not between different vertices.  In this diagram, the
main line is simply the entire horizontal arrow running along the bottom
of the diagram.
Thus, the only contribution to $s^{eff}$ will be the contribution
of Fig.~6.

For contributions to $t(0)$, we may define 2 main lines.  
One can follow the main line of the $G(l)$ or the main line of the $G^*(m)$.
These are the lines one follows if one starts on the line for $G(l)$ or
for $G^*(m)$ and follows through the various contractions as defined above.
These two lines join at some point in a noise contraction.

The second example of Fig.~8 has an  exceptional
contraction and may be ignored.  The main line starting with the $G(l)$
line leaving the circled $t$ vertex proceeds up and left, then turns right,
going straight across the diagram to the end.  Then it turn down and left,
up and left through an exceptional contraction into a $t$ vertex.  Then
it goes down and left into a noise contraction, where it ends.  The
main line starting with $G^*(m)$ starts at the circled $t$ vertex
and proceeds up and right until it terminates at the noise contraction.
Fig.~8b is very like Fig.~7b, except an additional $t$ vertex has
been added to the diagram.
If this vertex were removed, this diagram would be small for small
external momenta.  With the $t$ vertex on the diagram, the diagram
is very difficult to evaluate since to evaluate it requires a knowledge of
all $t(j,k)$, not just $t(\Lambda,\Lambda)$.  However, the diagram
is not only next order in the coupling constant, but also small due to
the various exponentials present, as may be verified. 
Any diagram for $t$ with such an exceptional contraction will have the same problems.  That is why we will ignore them.

If the main line of the $G^*(m)$ has lines leaving it which contract
against lines leaving the main line of $G(l)$ then the diagram will
again vanish due to the $\theta$ functions.  A contribution to $t$ cannot
have both external lines leaving the main line of $G^*(m)$.  Therefore
both lines must leave the main line of $G(l)$, and the line $G^*(m)$ 
cannot be dressed by any interaction vertices.
Again due to $\theta$-functions, the main line of $G(l)$ can only include 
one interaction vertex which has both external lines on it.  For
example, the diagram of Fig.~8d has external lines coming off different
interaction vertices and vanishes if the two lines are close in momentum.

All that remains is to justify the neglect of six-, and higher-,point functions,
such as could appear from the diagram in Fig.~8e.
If we wish to calculate a correlation function of four $G(j)$, and somewhere
in one diagrammatic calculation we have a six-point function, 
some of the lines leaving the six-point function must be contracted
against each other.  Then, somewhere in the diagram for the correlation
is a contribution to the four-point function which includes the six-point
function within it.  Therefore, the renormalization procedure would have
yielded this contribution to the four-point function as a change in $t$.
But, we have already, as outlined above, obtained all the contributions to
the change in $t$.  Therefore, there is no such diagram.

Finally, the lowest order contributions considered in the previous
subsection have a certain universality; considering only the diagrams of that
subsection, the nature of the fixed point does not depend on the initial
form of $t(j,k)$.  Higher order RG contributions will depend on the
initial form of $t(j,k)$.
\section{Fractal Dimension and Multifractal Exponents}
It is now possible to begin extracting exponents of the original
DLA model.  Different exponents correspond to different correlation
functions of this model; it will be the purpose of this section to determine
how to compute exponents from correlation functions.  This process
depends on the discussion of the adiabatic assumption and the assumption
used to introduce the cutoff into the continuum equation.  
From those assumptions an unambiguous means of determining exponents from
correlation functions is given.

In any actual simulation, there is an ultraviolet cutoff $\Lambda$ determined
by the ratio of macroscale to microscale.  
In the RG, a power law decay was found for 
$G(j)$.  Since the RG describes a fixed point in the scaling region,
within the RG itself this power law holds for arbitrarily large $j$.
Within an actual simulation this power law will fail at 
$j$ of order $\Lambda$, where $\Lambda$ is the cutoff of equation
(\ref{lf1}) resulting from the finite
size of the cluster in the simulation.
Thus, in the calculation of exponents that follows, although all correlation
functions are calculated using the rules of the RG and of perturbation
theory, the integrals over correlations functions that we will use must
be cutoff at momenta of order $\Lambda$, as will be done.

Since the size of the object follows a power law behavior given by
\begin{equation}
F_1\propto t^{1/D}
\end{equation}
where $D$ is the radius of gyration dimensions of the object, we have
\begin{equation}
\label{power}
\frac{d\log(F_1)}{dt}\propto{1/t}
\end{equation}
As a side point, strictly speaking this requires that $d F_1/dt$ can be
replaced by  the derivative of the average value of $F_1$,
but both numerical evidence and the RG flow of frequency resulting from
equation (\ref{rgw})
justify this assumption.
However, 
\begin{equation}
\label{llam}
dF_1/dt=\langle \lambda \rangle F_1
\end{equation}
where $\langle \lambda \rangle$ is 
defined to be the
average value of $\lambda$ over the unit circle at a given time.  
Equation (\ref{llam}) may be derived by using
equation (\ref{GPceq}) to calculate $dG_0/dt$ and
then equation (\ref{yo}) to relate this to $dF_1/dt$. 
Combining equations (\ref{power}),(\ref{llam}) we get
\begin{equation}
\label{elec}
\langle \lambda \rangle \propto 1/t
\end{equation}
Equation (\ref{elec}) is equivalent to the electrostatic scaling law first
derived by Halsey\cite{Halsey2}.  
In the continuum approximation, $\langle \lambda \rangle =\int dj \langle G(j) 
G^{*}(j)\rangle $.  
Here it must be understood that while in the perturbation theory this
expression is formally infinite, since $\langle G(j) G^{*} (j)\rangle
 =\delta(0) |G^2(j)|_{av}$,
in the above average we remove this factor of $\delta(0)$\cite{cluster}.
Calling $D$ the fractal dimension,
\begin{equation}
\label{sequence}
F_1^{-D} \propto 1/t \propto \langle \lambda\rangle \propto G_0^2 
\int^{\Lambda} dj j^{-2/5} \propto F_1^{-2}
\Lambda^{3/5} \propto F_1^{-2} F_{1}^{3/10}
\end{equation}
This gives the result that
\begin{equation}
D=2-1/2+1/5=1.7
\end{equation}

The first proportionality in equation (\ref{sequence}) followed from the radius of
gyration definition of the dimension.  The second followed from the
electrostatic scaling law.  The third followed from the expression for
$\langle \lambda\rangle $ in terms of $G$ and from the scaling of $G$ derived 
in the
RG.  The fourth followed from equation (\ref{yo}) and from doing the integral.
The fifth followed from
the functional dependence of $\Lambda$ on $F_1$ as given by
equation (\ref{lf1}).

This is the simplest way to derive the fractal dimension from the above work.
The calculation of the growth rate from equation (\ref{elec}) is essentially
a determination of the unrenormalized, unrescaled $s$ in equation (\ref{rseq}).
It may also be possible to repeat the same result by using the rescaling of $s$
under the RG to obtain the rescaling of the growth rate under a shift in
$\Lambda$. 

The multifractal exponents $\tau(q)$ are defined by 
\begin{equation}
\label{tqeq}
\tau(q)=\lim\limits_{l\rightarrow 0} \left(\log(\sum_i E^q(i)) /\log(l)\right)
\end{equation}
where the surface of the cluster
is covered with intervals of length $l$, and $E^q(i)$ is the integral
along the $i$th interval of the $q$th power of
the electric field.  Numerical calculations of
these exponents can be found in references \cite{Halsey1,yeah}.

One can try to compute higher multifractal exponents using the
RG (the work above amounts
to computing $\tau (3)$ and showing that $D=\tau(3)$).   
For example, the scaling of $\tau (5)$ 
can be determined by calculating the scaling of 
\begin{equation}
\label{t5}
\int^{\Lambda} dj \, dk \, dl \, dm \,
\langle G(j) G(k) G^{*}(l) G^{*}(m)\rangle \delta (j+k-l-m)
\end{equation}
against the upper cutoff $\Lambda$.
This is because the given integral is equal to the desired power of the
field integrated over the surface of the object.
If all the terms in this integral contributed with the same phase, the integral
would scale as $\Lambda^{-4/5+3}$.  Of course, the terms are independent
and this misestimates the exponent.  It is necessary to use the perturbation
theory to evaluate the 4 point correlation function.
The simplest possibility
is to use $G^f$ as an estimate for all the $G$ in equation (\ref{t5}).
The diagram for this is shown in Fig.~9a.
The only terms that would then contribute would be when a $G$ and a $G^{*}$ were at the
same momentum and the integral would scale as $\Lambda^{-4/5+2}$.  The different
scaling results from having a different number of momenta to integrate over.
Another possibility (this is analogous to a tree approximation for a 
scattering problem) is to substitute for the highest momentum $G$ in terms
of a $t$ vertex, leaving a six-point correlation function, and then take
all six $G,G^*$ to be $G^f$.  The scaling is then as $\Lambda^{-6/5+3}$.  
The diagram for this is shown in Fig.~9b. Since this scales more
strongly with $\Lambda$, it will be dominant in the limit needed to compute
$\tau(q)$.

The following is the rule for calculating multifractal
exponents:{\it Let $n$ be a positive integer.  Calculate the integral over
2n-point correlation functions defined by
\begin{equation}
\begin{array}{l}
\int^{\Lambda} dj_1 \int^{\Lambda} dj_2 . . . \int^{\Lambda}
 dj_n \int^{\Lambda} dk_1 \int^{\Lambda} dk_2 . . . \int^{\Lambda}
dk_n \\ \langle G(j_1) G(j_2) . . . G(j_n) G^*(k_1) G^*(k_2) . . . G^*(k_n)
\rangle
\delta(j_1+j_2+. . . +j_n - k_1 - k_2 - . . . - k_n)
\end{array}
\end{equation}
If this integral behaves, in the limit of $\Lambda$
taken to infinity, as $\Lambda^{a}$, where $a$ is some number, then
$\tau(2n+1)=(2n)-a/2$.}  The factor of $(2n)$ is the trivial scaling
that would result even for a nonfractal object; the factor of $a/2$ results
from the dependence on $\Lambda$ and from the square root dependence of
$\Lambda$ on length scale.

 In general, we can always find, for $\tau(q)$, a tree diagram that scales like
$\Lambda^{-2(q-2)/5+(q-2)}$.  Then, 
\begin{equation}
\tau(q)=(q-1)-1/2(-2(q-2)/5+(q-2))
\end{equation}
Alternately, another definition of exponents is
\begin{equation}
\label{dqd}
D_q=\tau(q)/(q-1)
\end{equation}
Then
\begin{equation}
D_q=\tau(q)/(q-1)=1-\frac{q(1/2-1/5)-1+2/5}{(q-1)}
\end{equation}
which is equivalent to
\begin{equation}
\label{dqeq}
D_q=\frac{0.7q-0.4}{q-1}
\end{equation}
\section{Comparison With Numerics and Discussion}
The theory is compared with numerics, and further tests of the theory are
proposed.

\subsection{Comparison With Numerics}
In previous work we found that the alternate formulation of DLA using
analytic functions\cite{us}
produces clusters with appearance and dimension similar
to those of clusters grown using the lattice formulation of DLA.
As far as we can tell, the two formulations are equivalent when $\alpha=2$.

The simplest comparison with numerics is the dimension itself.  1.7 is
very close to the accepted value of 1.71.

Equation (\ref{dqeq}) for higher multifractal exponents is the same
as the formula obtained with a wedge model by Halsey et. al.\cite{Halsey1},
except that
the wedge model left the quantities $0.7$ and $0.4$ as unknown constants
to be fitted to numerics.  They define
quantities $f$ and $\alpha$, the dimension of the set on which the wedges exist,
and the strength of the singularity (hopefully, the reuse of the symbol $\alpha$
will cause no confusion), and show that $D_q=\frac{\alpha q -f }{q-1}$. 
A numerical fit gave $\alpha=0.705$, $f=0.42$, while
a comparison with equation (\ref{dqeq}) gives $\alpha=0.7$, $f=0.4$.  
It is now known that
such a simple scaling law is not valid for large $q$\cite{yeah},
and in the original paper of Halsey et. al. it was suspected that
such a law would not hold.

The possible difference between theory and experiment here for
large $q$ should not be 
construed as a flaw in the presented work.  First, the above calculation
is only a lowest order calculation.  To higher orders, we may find a form
for $t(j,k)$ which has nontrivial behavior. This may alter the results 
from the tree approximation to the correlation function used to compute
the exponents.  Second, we may find interesting behavior if we consider
other contributions to the correlation exponents, beyond the tree diagrams 
used above.  Third, although the neglect of the appearance of 6-point
function was valid when considering the renormalization flow of $s$ and
$t$, as discussed in reference to Fig.~8e,,
such a neglect is not valid if one actually wishes 
to compute 6- and higher point
correlation functions.  Fortunately, such multi-point
interaction terms are captive
variables, in the sense that if one knows the behavior of $s,t$ under
RG flow one may systematically determine the higher interaction terms that
will appear.  Fourth, the above derivation of multifractal exponents
involved expressing the exponent in terms of correlation functions; this
is only possible for odd multifractal exponents.  Thus, in fact it is not
possible to say anything about even exponents in any simple fashion.

Additionally, there exist some difficulties in numerical calculation of higher
multifractal exponents.  According to the branched growth theory of DLA
\cite{Branch}, the
time required to compute higher exponents is superexponential in the order
of the desired exponent.  Thus, the exact values of the larger exponents may
not be precisely given by the numerical experiments.
This mathematical difficulty may be the source of the controversy
which appears to exist
between the different numerical calculations of these exponents.
For example, the value quoted for $\tau(3)$ by Ball et. al. is less
than 1.6, which is definitely at odds with the electrostatic scaling
law (believed to be exact from various numerical calculations),
and with other numerical calculations of this exponent.  The
electrostatic scaling law says that $\tau(3)=D=1.71$.  There also
exists controversy about the precise value of the dimension of
DLA, as mentioned in Ref. \cite{accepted}.
Thus, in fact, it is not clear exactly how large is the discrepancy between the
above results and the numerical results.

It is also of interest to numerically check the scaling of $G(j)$.  This was
check for two cases.  First, for a single cluster as described in
the next paragraph; second, for an ensemble of clusters as described in
the paragraph after that.

When $\int dj \,G(j) G^{*}(j) e^{-j/\Lambda}$ is plotted
against $\Lambda^{0.6}$, where in reality
the integral is a discrete sum, one expects to find a straight line behavior.
This is what is found, as shown in Fig.~10, except that for large $\Lambda$
the curve flattens out, since the numerical calculation only included a
finite number of terms.  Also for small $\Lambda$, the curve flattens out
at $F_1^{-2}$, which, in the long time limit, is vanishingly small compared
to the full integral. For the finite cluster size of our simulation,
$F_1^{-2}$ is not negligible.
The clusters here were grown using the conformal mapping
technique outlined previously. 
The coefficients of $G$ were computed with a numerical Fourier transform, by
mapping a large number of points on the surface of the circle (in fact,
slightly outside the circle, to improve numerical behavior) to the
surface of the aggregate (again, to slightly outside the surface
of the aggregate) and analytically calculating
the derivative of the mapping for each point.  This technique is not
very efficient for growing large clusters, at least as presently
implemented.  It requires ${\rm O}(N^2)$ time to compute $N$ growth steps, but
it is very easy to calculate coefficients of $G$ using this program. I 
only used aggregates of around 7000-10000 walkers.

As another check, 50 clusters of 6000 steps were simulated, and for
each cluster, the coefficients of $G$ were computed.  The squares of
these coefficients were scaled by $G(0)$, the overall inverse cluster size,
and then averaged together.  In Fig.~11 the mean squares
of $G(j)$ are plotted against $j$ on a log-log plot.
Numerical difficulties made it impossible to accurately extract
the slope in the scaling region.  This scaling region extends from
$j=5$ to $j=35$, or from $\log(j)=1.6$ to $\log(j)=3.5$.
Theory predicts that this slope is $2/5=0.4$.  The
numerical slope is between 0.3 and 0.5, using a least squares fit.  
The theory line is drawn onto
the graph.  As an additional check, another ensemble of clusters
was simulated, with a different $\lambda_0$ and a different number of
steps.  Within the scaling region of that simulation, the slope
of the mean squares behaved in the same fashion, and, additionally,
the mean-square of $G(j)$, after scaling by cluster size, for given
$j$, was the statistically the same for the two simulations.
\subsection{Discussion}
A theory has been presented based on the conformal nature of various
Laplacian growth processes.  A series of approximations were made that
produced a modified continuum equation of motion; it is hoped that
such an equation describes DLA, but even if it does not, it does
describe some form of nontrivial Laplacian growth.  A perturbation
theory was developed for this equation, and resummed.  To determine 
various terms in the perturbation theory, it was then necessary to
use a renormalization group calculation.  This has only been carried
out to lowest order.  It is a peculiar feature of this method that next
order calculations are vastly more difficult than lowest order calculations,
thus as yet there is no analytic calculation of higher order effects.
Finally, the assumptions leading to the modified model were reversed,
leading to calculations of quantities for DLA.

It would be worthwhile to look more closely at higher order corrections,
if not analytically, at least qualitatively, to see what may happen.  
To lowest order, $t(j-k)$ flows to an everywhere positive
function.
Using the lowest order $t$ to compute the effect of higher order
corrections 
will tend to lower the value for the dimension predicted
by this theory.  However, it is possible that in a more careful
next order calculation, the interaction $t(j-k)$ flows
to a function which is negative for large $(j-k)$, possibly increasing
the predicted dimension.  As mentioned above, next order effects depend
in some way upon the initial functional form of $t(j,k)$.  Lowest order
effects do not.

Unfortunately, it is not possible to carry out a stability analysis of
the fixed point of the lowest order RG.  All that may be said from the
above calculation is that if a fixed point exists, other than a
trivial fixed point for which $t$ goes to zero, then this fixed point is 
described by this RG.

It would also be worthwhile to try to extend this technique to
other Laplacian growth models, such as the dielectric breakdown model.
For the dielectric breakdown model\cite{DBE}, 
different values of $\eta$ correspond, in the continuum limit, to different
values of $\alpha$ in the conformal mapping model of section 2.1.  
The difference between the
dielectric breakdown model and our model is that, away from the
DLA case, our model uses the same growth probability over the
surface and varying walker size, while the dielectric breakdown
uses a varying growth probability and constant walker size.  Although this
alters the scale of the cluster in our case, we would expect the fractal
dimension of the cluster grown at a given $\alpha$ with the conformal
model to be the same as the dimension obtained from the dielectric
breakdown model with $\eta$ equal to $\alpha-1$.

One might naively try to apply the technique above to the case of $\alpha$ different
from 2, by replacing equation (\ref{Geq}) with the definition
$G=F_z^{-\alpha/2}$, and using an 
equation of motion similar to (\ref{rseq}), with different 
initial $t(j,k)$.  This would lead to physically absurd
results, and is in fact different from defining $G=F_z^{-1}$ and using
a modified equation of motion as described in the next paragraph.  
The difference is in how the noise term is
inserted.  It is important for the perturbation theory that products of $G^f$
may be pairwise decomposed, and this property means different things depending
on whether $G=F_z^{-1}$ or $G=F_z^{-\alpha/2}$.  
In the stochastic problem, each growth step produces a simple pole of $F_z^{-1}$
inside the unit circle; the angular coordinate of the pole is random, the
radial coordinate is determined by $\lambda$.  In the continuum limit
of section 2.5, the angular coordinate becomes the real value of $\theta$,
while the radial coordinate becomes the imaginary value of $\theta$;
the interior of the unit circle is replaced by the lower half plane.
One may show, using Cauchy's theorem, that randomly inserting 
{\it simple} poles produces a pairwise decomposition property for the
random noise in $F_z^{-1}$.
Therefore, the equation (\ref{Geq}) is the best definition of $G$.

One {\it can} handle the case of $\alpha$ different from 2 by using a
modified equation of motion, although this may be difficult if $\alpha$ is
not even.  One would modify equation (\ref{rseq}) by including
higher powers of $G$ in the interaction term.  For example, for
$\alpha=4$, the interaction term would be of the form 
\begin{equation}
\begin{array}{l}
(1/\mu) \int dk \, t(j,k) G(k) \int \int dl \, dm \, dn \, do \,
G(l) G^{*}(m) G(n) G^*(o) \\ \exp(-(l+m+n+o)/\Lambda)  \delta(k+l+n-m-j-o)
2 \theta(j-k) 
\end{array}
\end{equation}
This would probably be the most worthwhile test of the
calculations of this paper; although the calculation for $\alpha=4$ is
far more difficult than that for $\alpha=2$, it may still be tractable to
lowest order.  It would not be appropriate to attempt such a calculation
in this paper.  A few preliminary calculations show that one obtains
at least the physically correct result that the dimension of the $\alpha=4$
model is less than that of the $\alpha=2$ model; as yet the exact value at
$\alpha=4$ is not calculated\cite{cutoff}.
\section{Acknowledgements}
I would like to thank Leonid Levitov for
working with me on our previous paper on the conformal mapping model,
and for many useful
discussions that helped clarify the ideas outlined above.


\begin{thebibliography}{99}
\bibitem{cluster}
Actually, $\delta(0)$ is of order $1/\mu$, and so this formal
infinity is quite correct.  When integrating the electric field over
the circle, the result should be proportional to the size of the circle.
From the point of view of the continuum limit of section 2.5, 
this size is infinite, but in reality it is cut off at size $1/\mu$.
\bibitem{cutoff}
For $\alpha\neq2$, equation (\ref{roceq}) gets replaced by
${\rm Re}(F_t \frac{|F_z|}{z F_z})
=|F_z|^{1-\alpha} - \alpha |F_z|^{2-\alpha} \epsilon {\rm Re}(\frac{z F_{zz}}{F_z |F_z|})$
and so the balancing of electric field and surface tension of equation
(\ref{baleq}) is at the same value of $\epsilon$.  It is expected then, that
equation (\ref{lf1}) is that appropriate equation for any value of
$\alpha$.  We found from equation (\ref{roceq}) that $\delta=\epsilon$.
Since $\delta$ is the distance of the zero of $F_z$ to the boundary
of the unit circle, and $R\propto1/\delta^2$,
this suggests that equation (\ref{lf1}) may have
a geometrical explanation, as well as a dynamical explanation.
\end{thebibliography}
\end{document}